\documentstyle[12pt]{article}
\newcommand{\be}{\begin{equation}}
\newcommand{\ee}{\end{equation}}
\newcommand{\bea}{\begin{eqnarray}}
\newcommand{\eea}{\end{eqnarray}}
\newcommand{\ba}{\begin{array}}
\newcommand{\ea}{\end{array}}

\newcommand{\mathbb}[1]{\mathbf{#1}}

\begin{document}


\begin{center}
{\Large \bf  Eigenvalues and eigenstates of the
$s\ell_q(2)$-invariant Universal $R$-operator defined for cyclic
representations at roots of unity. }

\end{center}

\begin{center}

\vspace{1cm} {\large D.R. Karakhanyan}\\
{\em Yerevan Physics Institute, Armenia\\
(Alikhanian Brs. Str. 2, Yerevan 375036, Armenia)\\
E-mail: karakhan@lx2.yerphi.am}\\
\end{center}

\begin{abstract}
The $s\ell_q(2)$ representations are realized in the space of
polynomials for general and exceptional values of deformation
parameter $q$ and on finite set of theta-functions for cyclic
representation corresponding to $q^N=\pm 1$, which are a natural
extension of the polynomials. The complete set of eigenstates of
the Universal R-matrix are constructed and corresponding
eigenvalues are calculated.
\end{abstract}

\vspace{3mm}

\newpage


{\small \tableofcontents}
\renewcommand{\refname}{References.}
\renewcommand{\thefootnote}{\arabic{footnote}}
\setcounter{footnote}{0}
\setcounter{equation}{0}

\renewcommand{\theequation}{\thesection.\arabic{equation}}

\section{Introduction}
The representation of the symmetry groups in the functional space
seems to be interesting in the applications of theory of
integrable systems for construction of the Universal R-matrix
\cite{frt}, which intertwins two arbitrary representations,
because it provides a unique approach to the representations
regardless to their dimensions \cite{dkk}.\\
The representations of $q$-deformed (quantum) groups \cite{ks},
\cite{d} for positive real values of deformation parameter are, in
fact, the same as for non-deformed case. When $q$ takes arbitrary
complex values q-deformed universal enveloping algebra becomes
complex with non-unitary representations, which are less
interesting from the physical point of view. However, the
realization of the representation in the space of functions does
not differ from the case of positive real values of $q$.\\
The only exception is the case of complex roots of unity, which
has been considered at the early stage of studying of quantum
groups \cite{s}. The new type of representations (cyclic) appears
in this case. The unitary representations in this case exists and
appear in physical applications \cite{a}. The complete set of
irreducible representations of $s\ell(2)$ was presented in
\cite{ra}.\\
The eigenvalues and eigenvectors of Universal R-matrix were
calculated using functional representations for Heisenberg magnet
invariant with respect to $s\ell(2)$, $s\ell(2|1)$ and
$s\ell_q(2)$ symmetry groups. The Universal R-matrix in mentioned
cases was realized also as an integral operator \cite{dkk}.
\section{The case of isotropic (XXX) Heisenberg model}
It is reasonable to start the description of the proposed method
with discussion of more simple case of the isotropic, i.e.
$s\ell(2)$-invariant, Heisenberg spin chain. The integrability of
that model is based on the fundamental Yang-Baxter relation (YBE):
\be \label{ybe1}
R_{12}(u-v)R_{13}(u)R_{23}(v)=R_{23}(v)R_{13}(u)R_{12}(u-v), \ee
here operators $R_{ij}$ act on tensor product of two
representations of $s\ell(2)$ algebra: $V_i\otimes V_j$. In the
simplest case, when all of spaces $V_i$ correspond to the
fundamental (two-dimensional) representation, the solution to the
Yang-Baxter equation is given by the $R$-matrix of minimal
possible dimension $4\times 4$: \be \label{R1/2} R(u)=\left(
\begin{array}{cccc}
a&&&\\
&b&c&\\
&c&b&\\
&&&a
\end{array}\right),
\ee
where non-zero matrix elements are:
$$a=u+\eta,\quad b=u,\quad c=\eta.$$ Here $u$ is a spectral
parameter of the theory and $\eta$ is some model parameter, which
can be set equal to unity. The next in complexity solution is the
Lax operator, which corresponds to the case when YBE (\ref{ybe1})
is defined on an arbitrary and two fundamental spaces. \be
\label{lybe} R_{12}(u-v)L_1(u)L_2(v)=L_2(v)L_1(u)R(u-v). \ee
Originally the Lax operator appeared in Korteveg-de Vries equation
\cite{KdV} and some other famous problems \cite{lax}. In present
context it has form: \be \label{rll} L(u)=\left(
\begin{array}{cc}
u+\eta S&\eta S^-\\
\eta S^+&u-\eta S
\end{array}\right)=u+\eta\sigma_aS^a,
\ee where $\sigma_a$, $a=1,2,3$ are Pauli matrices. When spin of
the operators $S$, entering into the $L$-operator, is equal to
$1/2$, i.e. third space also is two-dimensional, they can be
identified with Pauli matrices: $S^a=\frac 12\sigma^a$ and that
definition coincides with (\ref{R1/2}). The relation (\ref{rll})
takes place due to the commutation relations of the $s\ell(2)$
algebra for operators $S^a$. \be \label{nda} [S,S^\pm]=\pm
S^\pm,\qquad [S^+,S^-]=2S. \ee An arbitrary representation of this
algebra can be realized in the space of polynomials of one real or
complex variable $x$. The general representation of $s\ell(2)$ is
specified by one parameter $\ell$, which is the spin of the
representation. The generators (\ref{nda}) can be realized as
differential operators: \be \label{ndo} S^-=\partial, \qquad
S=x\partial-\ell,\qquad S^+=2\ell x-x^2\partial . \ee In this
realization the lowest weight vector always exists. It is
annihilated by the lowering generator $S^-$ and is obviously given
by a constant function. The representation space is given then by
linear combinations of monomials: \be \label{per} \{x^n\}. \ee
This monomials appear after repeating action of rising generator
$S^+$ on the lowest weight vector. If parameter $\ell$ is
specified by half integer number: $2\ell=N$, this procedure is
ended on $N$-th step as it can be seen from definition
(\ref{ndo}). In this case the representation possesses also the
highest weight vector and is finite-dimensional $d=N+1$. Otherwise
the process of creating of new vectors is not finished and for
general complex values of $\ell$ the representation $V_\ell$ is
infinite-dimensional. The tensor product of two such
representations can be built in similar way in the space of
functions of two variables. The $s\ell(2)$ generators are given
then by the sum of corresponding generators: \be \label{ndo2}
S^-=\partial_1+\partial_2,\quad
S=x_1\partial_1+x_2\partial_2-\ell_1 -\ell_2, \ee
$$
S^+=2\ell_1x_1+2\ell_2x_2-x^2\partial_1-x^2\partial_2.
$$
The lowest weight vector $\varphi_0(x_1,x_2)$ is defined again as
a solution to the equation:
$$
S^-\varphi_0(x_1,x_2)=0,
$$
and is given by the monomials of translational invariant
difference: $(x_1-x_2)^n$. The whole space of tensor product
$V_{\ell_1}\otimes V_{\ell_2}$ is reduced to the space of
polynomials of variable $x_1$ with degree not more than
${2\ell_1}$ and of variable $x_2$ with the degree no more than
$2\ell_2$ if both parameters $2\ell_i$ are given by positive
integer numbers. Otherwise it is infinite-dimensional and given by
the linear space with basis: \be \label{fnm}
\varphi^m_n(x_1,x_2)=(S^+)^m\varphi_n(x_1,x_2),
\qquad\varphi_n(x_1,x_2)= (x_1-x_2)^n. \ee\\
The next in complexity solution is related to the case when one of
vector spaces in (\ref{lybe}) corresponds to the fundamental
representation of $s\ell(2)$ while two others are arbitrary ones.
Corresponding Yang-Baxter equation can be used as a defining
relation to determine the universal $R$-operator. Let third space
to be fundamental. Lax operators are $2\times 2$ matrices with
respect to the third space and are differential operators with
respect to the first and the second spaces, while the Universal
$R$-operator is inert (scalar) with respect to the third space and
is complicated pseudo-differential operator acting on the second
and third spaces. So YBE (\ref{lybe}) has to be understood as a
matrix relation $2\times 2$ both sides of which act on some
function, belonging to tensor product $V_1\otimes V_2$.\\
The authors of \cite{krs} used this approach to determine a
recurrent relation on the matrix elements of the universal
$R$-matrix in finite-dimensional case. To carry out the same
procedure in this case we shall use a manifest form of
$L$-operator and shall separate dependence on $u+v$ and $u-v$.
Then YBE takes the form:
$$
R_{12}(u)S^a=S^aR_{12}(u),
$$
$$
R_{12}(u)K(u)=\bar K(u)R_{12}(u),
$$
where we denoted
$$
K(u)=\left(
\begin{array}{cc}
\frac u2(S_2-S_1)+S_1S_2+S_1^-S_2^+&\frac u2(S_2^--S_1^-)
+S_1S_2^--S_1^-S_2\\
\frac u2(S_2^+-S_1^+)-S_1S_2^++S_1^+S_2&\frac u2(S_1-S_2)
+S_1S_2+S_1^+S_2^-
\end{array}\right),
$$
and
$$
\bar K(u)=\left(
\begin{array}{cc}
\frac u2(S_2-S_1)+S_1S_2+S_1^+S_2^-&\frac u2(S_2^--S_1^-)
+S_1^-S_2-S_1S_2^-\\
\frac u2(S_2^+-S_1^+)+S_1S_2^+-S_1^+S_2&\frac u2(S_1-S_2)
+S_1S_2+S_1^-S_2^+
\end{array}\right).
$$
The first relation expresses $s\ell(2)$-invariance of the
$R$-operator. This fact has the number of important sequences: the
action of the $R$-operator has to be defined on all vectors of
basis $(S^+)^m\varphi_n(x_1,x_2)$, $(m,n=0,1,2,3...)$. However due
to commutativity of $R$-operator with $S^+$ one can restrict
oneself only to consideration of the lowest weight vectors
$\varphi_n(x_1,x_2)$, because on vectors created from the lowest
weight ones by $S^+$ YBE will take place automatically. It follows
from the symmetry relations that $R$-operator commutes with
Casimir operator:
$$
{\bf{C}}=(S_1^a+S_2^a)^2,
$$
i.e. these operators have common set of eigenstates. In other
words, the lowest weight vectors and vectors created from them by
the action of rising generator $(S^+)^m\varphi_n(x_1,x_2)$ are
eigenstates of the $R$-operator with eigenvalues which independent
on $m$. Then, the matrices $K(u)$ and $\bar K(u)$ transform
covariantly with respect to the algebra $s\ell(2)$:
$$
[\sigma^a-S^a,K(u)]=0,\qquad [\sigma^a+S^a,\bar K(u)]=0.
$$
This relation means the linear dependence of corresponding
equations. In order to determine eigenvalues of the $R$-operator
it is enough to solve the simplest equation corresponding to the
right upper corner of matrices $K$ on lowest weight vectors. One
has:
$$
\!R(u)\!K^-(u)(x_1-x_2)^N\!\!=\!\!(\ell_1\!+\!\ell_2\!+\!1\!-\!N\!
-\! u)R(u)(x_1-x_2) ^{N-1}\!\!=\!\!\bar K^-\!R(u)(x_1-x_2)^N,
$$
and as vectors $(x_1-x_2)^N$ are eigenstates of the $R$-operator
one can deduce from here the following recurrent relation on
eigenvalues of the $R$-operator: \be \label{recr}
R_N(u)=-R_{N-1}(u)\frac{\ell_1+\ell_2+1-N-u}{\ell_1+\ell_2+1-N+u},
\ee
$$
R_N(u)=(-1)^NR_0(u)\prod_{n=1}^N\frac{\ell_1+\ell_2+1-n-u}
{\ell_1+\ell_2+1-n+u}.
$$
Using this recurrent relation one can restore explicit form of the
universal $R$-matrix for given value of parameters $\ell_1$ and
$\ell_2$. Some simple examples are given in Appendix.
\section{$s\ell_q(2)$ algebra.}
Arguments presented in previous section work almost without any
changes in other, more complicated cases. The case of usual so
called $q$-deformation, which corresponds to violation of
three-dimensional rotational symmetry of XXX Heisenberg model to
cylindrical symmetry of XXZ model is considered below. The case of
Heisenberg chain invariant with respect to the superalgebra
$s\ell(2|1)$ was considered in \cite{dkk}. The work devoted to
consideration of the model deformed by the dimensionful parameter,
which is more interesting from the physical point of view is in
progress \cite{kk}.\\
Mentioned symmetry violation in XXZ model described by
introduction a new parameter $\Delta=\cos\lambda$ into
Hamiltonian. This new model can be considered as deformation of
isotropic model and namely in that context the notion of quantum
group was appeared \cite{ft}. However there exists another point
of view \cite{blz}, according to which general integrable
two-dimensional quantum field theory at classic level is described
by Yang-Baxter equation corresponding to rational dependence of
the $R$-matrix on spectral parameter (analog of XXX chain), while
upon quantization the quantum fluctuations leads to the appearance
of quantum anomalies, describing violation, more correctly
deformation of some classic symmetries of the initial theory. It
is equivalent to deformation of the representation space
structure, i.e. eigenvalues and eigenstates of the physical
quantities such as anomalous dimensions.\\
So in this way, $q$- or quantum deformation of $s\ell(2)$ has two
aspects. The first one consists of deformation of algebra itself:
\be \label{qalg} [S,S^\pm]=\pm S^\pm,\qquad
[S^+,S^-]=\frac{q^{2S}-q^{-2S}}{q-q^{-1}}. \ee In the space of
polynomials $s\ell_q(2)$ generators can be realized through
differential operators as follows: \be \label{dqop} S^-=\frac 1x
\frac{q^{x\partial}-q^{-x\partial}}{q-q^{-1}}, \quad
S=x\partial-\ell,\quad
S^+=x\frac{q^{2\ell-x\partial}-q^{x\partial-2\ell}}{q-q^{-1}}. \ee
These operators more correctly should be called finite-difference
rather than differential taking into account that on test function
$f(x)$ these act according to formulae:
$$
q^{aS}f(x)=q^{-a\ell}f(q^ax),
$$
$$
S^-f(x)=\frac{f(qx)-f(q^{-1}x)}{x(q-q^{-1})},
$$
$$
S^+f(x)=x\frac{q^{2\ell}f(q^{-1}x)-q^{-2\ell}f(qx)}{q-q^{-1}}.
$$
One can pass from multiplicative action of these operators to the
additive one using simple change of variables $x=q^t$:
$$
q^S=q^{-\ell}e^{\partial_t}\qquad\qquad q^Sf(t)\equiv
q^{-\ell}f(t+1)
$$
\be \label{fdot}
S^-=q^{-t}\frac{e^{\partial_t}-e^{-\partial_t}}{q-q^{-1}},\qquad
S^-f(t)\equiv q^{-t}\frac{f(t+1)-f(t-1)}{q-q^{-1}} \ee
$$
S^+=q^t\frac{q^{2\ell}e^{-\partial_t}-q^{-2\ell}e^{\partial_t}}{q-q^{-1}},
\qquad S^+f(t)\equiv
q^t\frac{q^{2\ell}f(t-1)-q^{-2\ell}f(t+1)}{q-q^{-1}}.
$$
In this form the origin of name finite-difference operator becomes
absolutely clear.\\
As for the case of non-deformed symmetry the representation of the
algebra is realized in the space of polynomials. This argument was
crucial for the choice (\ref{dqop}) of spin generators. Indeed,
the operator of finite-difference derivative $S^-$ annihilates a
constant function, which can be chosen as the lowest weight
vector. Then acting repeatedly by rising generator one creates
higher powers of variable $x$ and that process is ended only if
parameter $2\ell$ is given by positive integer. In this case
representation possesses also highest weight vector $x^{2\ell}$
and is finite-dimensional. Otherwise it is infinite-dimensional.\\
Algebra (\ref{qalg}) possesses the central element, Casimir
operator. (In this section we consider the case of general complex
values of deformation parameter $q$ different from the roots of
unity.)\\
\be \label{co}
{\bf{C}}=S^+S^-+[S]_q[S-1]_q=S^-S^++[S]_q[S+1]_qf(x). \ee here
notation $[N]_q$ is used for $q$-deformed numbers: $[N]_q=
\frac{q^N-q^{-N}}{q-q^{-1}}$. It is easy to see that in above
mentioned realization of spin operators (\ref{dqop}) the Casimir
is proportional to unity:
$$
{\bf{C}}f(x)=[\ell]_q[\ell+1]_q.
$$
So, for the general values of deformation parameter $q$ the
finite-dimensional representations are given by the same spaces
${\mathbb{C}}^{2\ell+1}$ as in non-deformed case. The case of
arbitrary complex $\ell$ corresponds to the infinite-dimensional
space of polynomials. The case $|q|=1$ requires more careful
analysis, in particular for $q^N=\pm 1$, the new type
representations appear. Those representations called cyclic ones
and have no classic (non-deformed) analog. In this section we
restrict ourselves to case $|q|\neq 0$. Such kind of deformation
reduces complex plane to a strip. It becomes obvious by
considering spin operators in terms of variable $t$
(\ref{fdot}).\\
In order to construct the representation with lowest (highest)
weight vector it is necessary to solve the following equation:
$$
S^-\varphi_0=0,\qquad \left(S^+\varphi_0=0\right),
$$
for the lowest (highest) vector. One has:
$$
S^-\varphi_0(t)=\frac{q^{-t}}{q-q^{-1}}\left[\varphi_0(t+1)-\varphi_0
(t-1)\right]=0, \qquad \varphi_0(t+1)=\varphi_0(t-1).
$$
In other words the lowest weight vector is given by functions of the form:
$$
\varphi_0(t)=\sum_ka_ke^{i\pi kt},
$$
with arbitrary complex coefficients $a_k$. Then the lowest weight
vector creates the whole representation space :\\
\be
\{\varphi_n(t)\}=\{(S^+)^n\varphi_0(t)\}, \ee
$$
\varphi_n(t)=(S^+)^n\varphi_0(t)=[2\ell]_q[2\ell-1]_q\ldots
[2\ell-n+1]_q\sum_k(-1)^{nk}a_ke^{(i\pi k+n\log q)t}.
$$
The lowest weight vectors $\varphi_0^{(k)}(t)\equiv a_ke^{i\pi
kt}$ appears to be eigenstates of operator of third projection of
spin:
$$
q^S\varphi_0^{(0)}(t)=(-1)^kq^{-\ell}\varphi_0^{(k)}(t).
$$
These create the equivalent representations:
$\{\varphi_0^{(k)}(t)\}= \{q^{nt}e^{i\pi kt}\}$. However remaining
in the frame of the strip $\arg q \in [0, 2\pi ]$ one can ignore
that difference and consider the representation realized in terms
of variable $x$.\\
Another aspect of transition from $s\ell(2)$ to $s\ell_q(2)$ is
related to deformation of tensor product. The simple rule of spin
addition $S^a=S_1^a+S_2^a$ is not working, because the sum of
generators does not satisfy to the algebra {\ref{qalg}). The
simple scaling arguments lead to the following modification:
$$
S=S_1+S_2,
$$
(this is because the symmetry with respect to rotations around the
$z$-axis preserves)
$$
S^+=q^{aS_1}S_2^++q^{bS_2}S_1^+,\qquad
S^-=q^{cS_1}S_2^-+q^{dS_2}S_1^-.
$$
Then compound generators satisfy to commutation relations
(\ref{qalg}) upon conditions $d=-a$, $c=-b$ and $a-b=\pm 2$. In
this way two different {\it{co-product}} can be introduced
(remaining parameter $a$ can be absorbed by redefinition of
operators):
$$
S^+=S^+_1q^{S_2}+ q^{-S_1}S^+_2,
$$
\be \label{S} \Delta:\quad\qquad S=S_1+S_2,\qquad\qquad \ee
$$
S^-=S^-_1q^{S_2}+ q^{-S_1}S^-_2,
$$
and
$$
\bar S^+=S^+_1q^{-S_2}+ q^{S_1}S^+_2,
$$
\be \label{barS} \bar\Delta:\quad\qquad\bar S=S_1+S_2,\qquad\qquad
\ee
$$
\bar S^-=S^-_1q^{-S_2}+ q^{S_1}S^-_2.
$$
Note that introduced co-products do not preserve symmetry with
respect to exchange $q\leftrightarrow q^{-1}$. It means that
building the tensor product using of them one will deal with two
equivalent but different realizations of the same space $V$ ($\bar
V$), corresponding to $\Delta$ ($\bar\Delta$). The spaces $V$ and
$\bar V$ built from linear combinations of vectors:
$(S^+)^m\varphi_N(x_1,x_2|q)\equiv\varphi^m_N (x_1,x_2|q)$ and
$(\bar S^+)^m\bar\varphi_N(x_1,x_2|q)\equiv\bar\varphi
^m_N(x_1,x_2|q)$ correspondingly (so called {\it{Drinfel'd quantum
double}}) \cite{d}. Here we introduce the lowest weight vectors
\be \label{lwv}
\varphi_N(x_1,x_2|q)=\prod_{n=1}^N(q^{\ell_1+1-n}x_1-x_2q^{n-1-\ell_2}),
\ee which correspond to solutions to the equations:
$$
S^-\varphi_N(x_1,x_2|q)=0,
$$
and similarly for over barred quantities:
$\bar\varphi_N(x_1,x_2|q)=\varphi_N(x_1,x_2|q^{-1})$. The vectors
$\varphi^m_N(x_1,x_2|q)$ ($\bar\varphi^m_N(x_1,x_2|q)$), form
basis in the space $V$ ($\bar V$) and are eigenstates of Casimir
operators
$$
{\bf{C^{(2)}}}=q^{S_2-S_1+1}S_1^+S^-_2+q^{S_2-S_1-1}
S_1^-S_2^+-(q-q^{-1})^{-2}\left((q+q^{-1})\left(1+q^{2S_2-2S_1}\right)
\right)+
$$
$$
q^{2S_2}[\ell_1]_q[\ell_1+1]_q+q^{-2S_1}[\ell_2]_q[\ell_2+1]_q
$$
and
$$
{\bf{\bar C^{(2)}}}=\left(q^{S_1-S_2-1}S_1^+S^-_2+q^{S_1-S_2+1}
S_1^-S_2^+\right)-(q-q^{-1})^{-2}\left((q+q^{-1})\left(1+q^{2S_1-2S_2}
\right)\right)+
$$
$$
q^{-2S_2}[\ell_1]_q[\ell_1+1]_q+q^{2S_1}[\ell_2]_q[\ell_2+1]_q,
$$
correspondingly
$$
{\bf{C}}(S^+)^m\varphi_N(x_1,x_2|q)=[N-\ell_1-\ell_2]_q
[N-\ell_1-\ell_2-1]_q(S^+)^m\varphi_N(x_1,x_2|q).
$$
The same is true for over barred quantities. Note that like in
non-deformed case the eigenstates are degenerate over index $m$.
\section{Yang-Baxter relation}
The solution to the YB relation for XXZ spin $\frac 12$ chain has
the same form as for isotropic chain (\ref{R1/2}) but with
trigonometric dependence on spectral parameter $u$:
$$a=q^{u+1}-q^{-u-1},\quad b=q^u-q^{-u},\quad
c=q-q^{-1},\quad$$. The deformation parameter $q=e^{i\lambda}$
enters here through anisotropy parameter of XXZ chain:
($\cos\lambda=\Delta$). We use notations of article \cite{f}. The
fundamental representation is again given by the Pauli matrices as
for $2\times 2$ matrices the relations (\ref{qalg}) coincide with
non-deformed ones. The Lax operator of XXZ model is given by the
expression \be \label{L} L_{\ell,a}(u)=\left(
\begin{array}{cc}
q^{u+S}-q^{-u-S}&(q-q^{-1})S^-\\
(q-q^{-1})S^+&q^{u-S}-q^{-u+S}
\end{array}\right).
\ee The Yang-Baxter relation for two fundamental and one arbitrary
representation take place due to commutation relations
(\ref{qalg}) for spin operators. When $q\to 1$ these relations go
to the non-deformed (\ref{nda}).\\
In order to determine the universal $R$-operator we consider YBE
for spins $\ell_1$, $\ell_2$ and $\frac 12$: \be \label{ybe}
\!R_{\ell_1,\ell_2}(u\!-\!v)\!\! \left(\begin{array}{cc}\!\!\!
q^{u+S_1}\!-\!q^{-u-S_1}&(q-q^{-1})S_1^-\\
(q-q^{-1})S_1^+&q^{u-S_1}\!-\!q^{-u+S_1}
\end{array}\!\!\!\right)\!\!\!\left(\!\!\!\begin{array}{cc}
q^{v+S_2}\!-\!q^{-v-S_2}&(q-q^{-1})S_2^-\\
(q-q^{-1})S_2^+&q^{v-S_2}\!-\!q^{-v+S_2}
\end{array}\!\!\!\right)\!\!=
\ee
$$
\!\left(\!\!\!\begin{array}{cc}
q^{v+S_2}\!-\!q^{-v-S_2}&(q-q^{-1})S_2^-\\
(q-q^{-1})S_2^+&q^{v-S_2}\!-\!q^{-v+S_2}
\end{array}\!\!\!\right)\!\!\!\left(\!\!\!\begin{array}{cc}
q^{u+S_1}\!-\!q^{-u-S_1}&(q-q^{-1})S_1^-\\
(q-q^{-1})S_1^+&q^{u-S_1}\!-\!q^{-u+S_1}
\end{array}\!\!\!\right)\!\!R_{\ell_1,\ell_2}
(u\!-\!v).
$$
As mentioned above, the main intertwining property of the
$R$-operator consists in that it intertwines two different,
corresponding to co-products $\Delta$ and $\bar\Delta$
realizations of the spaces of representations $V$ and $\bar V$:\\
\be \label{intertwin} R(u)\Delta=\bar\Delta R(u). \ee Separating
in YBE dependence on $u-v$ and $u+v$ one obtains that it
equivalent to the following set of equations: \be \label{pm2}
[R_{\ell_1,\ell_2}(u),q^{S_1+S_2}]=0, \ee \be \label{+1-}
R(u)\left(q^{\frac u2+S_1}S_2^-+q^{-\frac u2-S_2}S_1^-\right)=
\left(q^{\frac u2-S_1}S_2^-+q^{-\frac u2+S_2}S_1^-\right)R(u), \ee
\be \label{+1+} R(u)\left(q^{\frac u2-S_1}S_2^++q^{-\frac
u2+S_2}S_1^+\right)= \left(q^{\frac u2+S_1}S_2^++q^{-\frac
u2-S_2}S_1^+\right)R(u), \ee \be \label{-1-} R(u)\left(q^{\frac
u2+S_2}S_1^-+q^{-\frac u2-S_1}S_2^-\right)= \left(q^{\frac
u2-S_2}S_1^-+q^{-\frac u2+S_1}S_2^-\right)R(u), \ee \be
\label{-1+} R(u)\left(q^{\frac u2-S_2}S_1^++q^{-\frac
u2+S_1}S_2^+\right)= \left(q^{\frac u2+S_2}S_1^++q^{-\frac
u2-S_1}S_2^+\right)R(u), \ee ?
$$
R(u)\left(q^{u+S_1-S_2}+
q^{-u+S_2-S_1}-(q-q^{-1})^2S_1^-S_2^+\right)=
$$
\be
\label{0-} \left(q^{u+S_1-S_2}+q^{-u+S_2-S_1}
-(q-q^{-1})^2S_1^+S_2^-\right)R(u),
\ee
$$
R(u)\left(q^{u+S_2-S_1}+q^{-u+S_1-S_2}-
(q-q^{-1})^2S_1^+S_2^-\right)=
$$
\be \label{0+} \left(q^{u+S_2-S_1}+q^{-u+S_1-S_2}
-(q-q^{-1})^2S_1^-S_2^+\right)R(u). \ee One deduces from this set
that it compatible with general {\it{unitarity}} property of
$R$-operator:
$$
R^{-1}(u)=R(-u).
$$
Indeed, multiplying (\ref{-1-}) by $R^{-1}(u)$ from the right and
from the left one sees that operator $R^{-1}(u)$ satisfies to the
relation (\ref{+1-}) with exchange $u\leftrightarrow -u$, i.e.
$R^{-1}(u)$ and $R(-u)$ satisfy to the same equations (the same is
true for eqs. (\ref{-1+}) and (\ref{+1+}), (\ref{0-}) and
(\ref{0+})).\\
Moreover, being physically observable quantity $R$-matrix has to
be single-valued function of $q$, i.e. has to be periodic function
of $\log q$. In other words $R$-matrix has to be inert to the
choice of a strip $\arg q$ on complex plane.\\
It follows from the YBE (\ref{pm2}-\ref{0+}) that algebra
$s\ell_q(2)$ is realized by the spin operators which are in
addition to definitions (\ref{S}) and (\ref{barS}) twisted by
spectral parameter:
$$
S^-_u=q^{\frac u2+S_2}S^-_1+q^{-\frac u2-S_1}S_2^-,
$$
\be \label{su} S^+_u=q^{-\frac u2+S_2}S^+_1+q^{\frac u2-S_1}S_2^+,
\ee
$$
S_u=S_1+S_2,
$$
and
$$
\bar S^-_u=q^{-\frac u2-S_2}S^-_1+q^{\frac u2+S_1}S_2^-,
$$
\be \label{barsu} \bar S^+_u=q^{\frac u2-S_2}S^+_1+q^{-\frac
u2+S_1}S_2^+, \ee
$$
\bar S_u=S_1+S_2.
$$
This extension is compatible with $s\ell_q(2)$-invariance of
co-products. The Yang-Baxter equations in terms of these operators
take more compact form:
$$
[R(u),S]=0,
$$
$$
R(u)\bar S_u^-=S_{-u}^-R(u),
$$
$$
R(u)\bar S_u^+=S_{-u}^+R(u),
$$
$$
R(u)S_u^+=\bar S_{-u}^+R(u),
$$
$$
R(u)S_u^-=\bar S_{-u}^-R(u).
$$
It allows to express the main intertwining property of
$R$-operator as follows: \be \label{interu}
R(u)\Delta_u=\bar\Delta_{-u}R(u) \ee which implies
$$
{\bf{C}}_{-u}R(u)=R(u){\bf{\bar C}}_u,\qquad\quad {\bf{\bar
C}}_{-u}R(u)=R(u){\bf{C}}_u.
$$
These relations express $s\ell_q(2)$-invariance of $R$-operator,
the eigenproblem for which can be formulated as follows: \be
\label{eigenp}
R(u)\bar\varphi^m_N(x_1,x_2|q,u)=R_N\varphi^m_N(x_1,x_2|q,-u), \ee
$$
R(u)\varphi^m_N(x_1,x_2|q,u)=\tilde
R_N\bar\varphi^m_N(x_1,x_2|q,-u),
$$
where vectors $\varphi$ are the eigenfunctions of Casimir
operators constructed from spin operators (\ref{su}) and
(\ref{barsu}). In other words, operator $R(u)$ acts on
{\it{Drinfel'd quantum double}} $V\oplus\bar V$ in skew-symmetric
way:
$$
R(u)\left(
\begin{array}{cc}
\varphi^m_N\\
\bar\varphi^m_N
\end{array}\right)=
\left(
\begin{array}{cc}
0&\tilde R_N\\
R_N&0
\end{array}
\right) \left(
\begin{array}{cc}
\varphi^m_N\\
\bar\varphi^m_N
\end{array}
\right).
$$
If one arranges equations: \be \label{s0}
Rq^{\pm(S_1+S_2)}=q^{\pm(S_1+S_2)}R, \ee \be \label{-}
S^-_{-u}R=R\bar S_u^-, \ee \be \label{+} S_{-u}^+R=R\bar S_u^+,
\ee to be the symmetry relations of the $R$-operator and denoting
remaining equations as follows: \be \label{k-} RK^-=\bar K^-R, \ee
\be \label{k+} RK^+=\bar K^+R, \ee \be \label{k+-} RK^{+-}=\bar
K^{+-}R, \ee \be \label{k-+} RK^{-+}=\bar K^{-+}R, \ee
$$
K^-=S_u^-,\quad \bar K^-=\bar S_{-u}^-,\quad K^+=S_u^+,\quad\bar
K^+=\bar S_{-u}^+,
$$
$$
K^{+-}=q^{u+S_1-S_2}+q^{-u+S_2-S_1}-(q-q^{-1})^2S^-_1S_2^+,
$$
$$
\bar K^{+-}=q^{u+S_1-S_2}+q^{-u+S_2-S_1}-(q-q^{-1})^2S^+_1S_2^-,
$$
$$
K^{-+}=q^{u+S_2-S_1}+q^{-u+S_1-S_2}-(q-q^{-1})^2S^+_1S_2^-,
$$
$$
\bar K^{-+}=q^{u+S_2-S_1}+q^{-u+S_1-S_2}-(q-q^{-1})^2S^-_1S_2^+.
$$
one obtains:
$$
[\bar S^-_u,K^-]=0=[S^-_{-u},\bar K^-],\qquad [\bar
S^+_u,K^+]=0=[S^+_{-u},\bar K^+],
$$
$$
[\bar S^-_u,K^+]=(q-q^{-1})^{-1}\left(q^{-S_1-S_2}K^{+-}-
q^{S_1+S_2}K^{-+}\right),
$$
$$
[S^-_{-u},\bar K^+]=(q-q^{-1})^{-1}\left(-q^{-S_1-S_2}\bar K^{+-}+
q^{S_1+S_2}\bar K^{-+}\right),
$$
$$
[\bar S^+_u,K^-]=(q-q^{-1})^{-1}\left(-q^{-S_1-S_2}K^{-+}+
q^{S_1+S_2}K^{+-}\right),
$$
$$
[S^-_{-u},\bar K^+]=(q-q^{-1})^{-1}\left(q^{-S_1-S_2}\bar K^{-+}-
q^{S_1+S_2}\bar K^{+-}\right).
$$
So like in non-deformed case quantities $K$ ($\bar K$'s) are
linearly dependent and one can restrict oneself to consideration
of symmetry relations (\ref{s0}-\ref{+}) and the simplest
(\ref{k-}) equation among remaining ones.\\
Let us turn now to the "eigenproblem" of the $R$-operator in sense
of definition (\ref{eigenp}) or that is the same, to the problem
of finding eigenstates of Casimir operators ${\cal{C}}_u$ and
$\bar{\cal{C}}_u$. It follows from the relation (\ref{interu})
that mentioned eigenvectors have the form: \be \label{eigenv}
\varphi^m_N(x_1,x_2|q,u)=(S^+_u)^m\varphi_N(x_1,x_2|q,u), \ee and
$$
\bar\varphi^m_N(x_1,x_2|q,u)=\varphi^m_N(x_1,x_2|q^{-1},u),
$$
where the lowest weight vectors $\varphi_N(x_1,x_2|q,u)$ are
defined as solutions to the equation
$$
S^-_u\varphi(x_1,x_2)=0,
$$
in form of polynomials of degree not more than $N$ over variables
$x_1$ and $x_2$: \be \label{eigenf}
\varphi_N(x_1,x_2|q,u)=\prod_{n=1}^N(q^{\ell_1+1-n-\frac
u2}x_1-q^{\frac u2+n-1-\ell_2}), \ee
$$
\bar\varphi_N(x_1,x_2|q,u)=\varphi_N(x_1,x_2|q^{-1},u)=
\prod_{n=1}^N(q^{-\ell_1-1+n+\frac u2}x_1-q^{-\frac
u2-n+1+\ell_2}).
$$
It is easy to see that functions $\bar\varphi(x_1,x_2|q,u)$ differ
from $\varphi_N(x_1,x_2|q,u)$ by formal exchange $(x_1,\ell_1)$
$\leftrightarrow$ $(x_2,\ell_2)$. It can be shown also
$$
\!\!S_u^-\!\!\left(\!\!\!
\begin{array}{cc}
\varphi_N(x_1,x_2|q,\!u)\\
\bar\varphi_N(x_1,x_2|q,\!u)
\end{array}\!\!\!\right)\!\!=\!\!
\left(\!\!\!
\begin{array}{cc}
0\\
-(q\!-\!q^{-1})[N]_q[\ell_1\!\!+\!\!\ell_2\!\!+1\!\!-\!\!N\!\!-\!\!u]_q
\bar\varphi_{N-1}(x_1,x_2|q,\!-u)
\end{array}\!\!\!
\right)\!\!,
$$
$$
S_{u}^-\left(\!\!\!
\begin{array}{cc}
\varphi_N(x_1,x_2|q,\!-u)\\
\bar\varphi_N(x_1,x_2|q,\!-u)
\end{array}\!\!\!\right)\!\!=\!\!
\left(\!\!\!
\begin{array}{cc}
(q\!-\!q^{-1})[u]_q[N]_qq^{\ell_1-\ell_2}\varphi_{N-1}(q^{-1}x_1,qx_2|q,\!-u)\\
(q\!-\!q^{-1})[N]_q[N\!-\!1\!-\!\ell_1\!-\!\ell_2]_q
\bar\varphi_{N-1}(x_1,x_2|q,\!-u)
\end{array}\!\!\!\right)\!\!,
$$
Another relations appearing in eigen-problem (\ref{eigenp}) can be
obtained from this one by exchange $q\leftrightarrow q^{-1}$ or
$u\leftrightarrow -u$. Taking into account that \be
\label{eigenprob} R(u)\left(
\begin{array}{cc}
\varphi_N(x_1,x_2|q,u)\\
\bar\varphi_N(x_1,x_2|q,u)
\end{array}\right)=
\left(
\begin{array}{cc}
0&\tilde R_N\\
R_N&0
\end{array}
\right) \left(
\begin{array}{cc}
\varphi_N(x_1,x_2|q,-u)\\
\bar\varphi_N(x_1,x_2|q,-u)
\end{array}
\right), \ee and acting by the relation (\ref{k-}) to
$\varphi_N(x_1,x_2|q,u)$ one obtains zero in both sides, while
acting to $\bar\varphi_N(x_1,x_2|q,u)$ one obtains desirable
recurrent relation on eigenvalues $R_N$'s: \be \label{recrel}
R_N=-R_{N-1}
\frac{[\ell_1+\ell_2+1-N-u]_q}{[\ell_1+\ell_2+1-N+u]_q} \ee
multiplied on $\varphi_{N-1}(x_1,x_2|q,-u)$. Then acting by the
relation (\ref{-}) to $\bar\varphi_N(x_1,x_2|q,u)$ one obtains
identity $0=0$ and acting to $\varphi_N(x_1,x_2|q,u)$ one obtains
another relation:
$$
\tilde R_N=-\tilde R_{N-1}
\frac{[\ell_1+\ell_2+1-N-u]_q}{[\ell_1+\ell_2+1-N+u]_q},
$$
multiplied on $\bar\varphi_{N-1}(x_1,x_2|q,-u)$. Notice that due
to the relation $\varphi_0(x_1,x_2|q,u)=\bar\varphi_0(x_1,x_2|q,u)
=1$ the initial conditions of both recurrent relations are the
same too $R_0=\tilde R_0$ so both quantities coincide for all
values $n$
$$
R_{N}=\tilde R_{N}=(-1)^NR_0\prod_{n=1}^N
\frac{[\ell_1+\ell_2+1-n-u]_q}{[\ell_1+\ell_2+1-n+u]_q}.
$$
That means that eigenvalues of universal $R$-operator are inert
with respect to exchange $q\to q^{-1}$. Notice also that
(\ref{recrel}) differs from corresponding non-deformed expression
consist of the replacement of usual numbers to quantum ones.
\section{Universal $R$-operator for the case of exceptional values of $q$.}
In this section we consider the case of exceptional values of
deformation parameter, corresponding to complex roots of unity of
degree $N$. We set for definiteness $q=\exp(2\pi i/N)$ where $N$
is odd number.\\
In this case the center of algebra $s\ell_q(2)$ is extended and
three more Casimir operator appear: $q^NS$ and $(S^\pm)^N$. Indeed
the relation (\ref{qalg}) in this case gives:
$$
(q^S)^NS^\pm=q^{\pm N}S^\pm q^{NS}=S^\pm(q^S)^N,
$$
$$
q^S(S^-)^N=q^N(S^-)^Nq^S=(S^-)^Nq^S,
$$
$$
[S^+,(S^-)^N]=\sum_{n=0}^{N-1}(S^-)^{N-n-1}[S^+,S^-](S^-)^n=
$$
$$
\frac{1}{q-q^{-1}}\sum_{n=0}^{N-1}(S^-)^{N-n-1}\left(
q^{-2n}(S^-)^nq^{2S}-q^{2n}(S^-)^nq^{-2S}\right)=
$$
$$
\frac{(S^-)^{N-1}}{q-q^{-1}}\left(q^{2S}\sum_{n=0}^{N-1}q^{-2n}-
q^{-2S}\sum_{n=0}^{N-1}q^{2n}\right)=0.
$$
Here we took into account that for mentioned values of $q$ the
both sums turn to be zero due to the formula of geometric
progression: $\sum_{n=0}^{N-1}q^{2n}=\frac{q^{2N}-1}{q^2-1}=0$.
Similarly one has:
$$
[(S^+)^N,S^-]=0,\qquad (S^+)^Nq^S=q^S(S^+)^N.
$$
That means that for mentioned operators the relations: \be
\label{ce} (S^\pm)^N=\alpha^\pm{\mathbb{I}},\qquad
q^{NS}=q^{N\gamma}{\mathbb{I}}, \ee must take place on {\it{all}}
vectors of representation. So the functions which form the
representation specified by parameters $\ell$, $\gamma$ and
$\alpha^\pm$ are determined as the solutions to the set of
equations: \be (S^\pm)^Nf(t)=\alpha^\pm f(t), \ee
$$
q^{NS}f(t)=q^{N\gamma}f(t).
$$
So for exceptional values of deformation parameter the center of
$s\ell_q(2)$ algebra is extended, i.e. the representations are
specified by more number of parameters related to the eigenvalues
of new Casimir operators, which are constrained by one algebraic
condition. It was shown in famous work of Kac and De Concini
\cite{dk} that in general case the representations of the
$s\ell_q(2)$ algebra with deformation parameter $q$ is given by
complex root of unity are specified by three parameters. Indeed
definition (\ref{dqop}) is not most general one. Commutation
relations (\ref{qalg}) can be realized by operators: \be
\label{dq1op}\!\! S^-\!=\!q^{-\frac\lambda 2}\frac 1x
\frac{q^{x\partial-\beta}\!-\!q^{\beta-x\partial}}{q-q^{-1}}\!,
\;\; q^S\!=\!q^{-\frac 12(\alpha+\beta)}q^{x\partial}\!,\;\;
S^+\!=\!xq^{\frac\lambda 2
}\frac{q^{\alpha-x\partial}\!-\!q^{x\partial-\alpha}}{q-q^{-1}}.
\ee However in the case of general values of $q$ using of this
expression is meaningless because algebra possesses only one free
parameter $2\ell=\alpha+\beta$ related to the unique Casimir
operator.\\
It follows from the definition (\ref{dq1op}) that Casimir operator
$q^{NS}$ acts trivially in functional space:
$$
q^{NS}f(x)=q^{-\frac{\alpha+\beta}{2}}f(q^Nx)
=q^{-\frac{\alpha+\beta}{2}}f(x),
$$
and is multiple to unity operator i.e. does not provide any
restriction on the form of functions which form representation.\\
The condition $\alpha^+(\alpha^-)=0$ means that operator
$S^+(S^-)$ is nilpotent, i.e. there exists vector, which
annihilated by that operator and representation possesses highest
(lowest) weight vector. Vise versa, if one supposes that highest
(lowest) weight vector does not exist then it would contradict to
nilpotency of operator $S^+(S^-)$. So the representation
corresponding to the case $\alpha^+(\alpha^-)=0$ can be obtained
by setting parameter $q$ equal to the root of unity in already
considered case of representations with highest (lowest) weight
vector for general values of $q$. So below we suppose that both
parameters $\alpha^\pm$ differ from zero. Consider first the
equation for operator $(S^-)^N$ (equation for $(S^+)^N$ can be
considered similarly). Commuting multipliers $q^{-t}=x^{-1}$ to
the left one obtains:
$$
(S^-)^Nf(x)=q^{-Nt}\prod_{n=0}^{N-1}[x\partial+n]_qf(x)=\alpha^-f(x).
$$
Consider the function:
\be
\label{phi}
\Phi_N(\alpha)\equiv\prod_{n=0}^{N-1}[\alpha+n]_q.
\ee
From the simple identity
$$
[\alpha+kN]_q=[\alpha]_q,\qquad k\in {\mathbb{Z}}
$$
one can establish periodicity property:
$$
\Phi_N(\alpha+1)=\Phi_N(\alpha).
$$
It follows from this property that in expansion of the product
(\ref{phi}) (for $N$ odd) to the sum only terms multiple to
$q^{\pm N\alpha}$ can survive \be
\Phi_N(\alpha)=(q-q^{-1})^{-N}(q^{\alpha N}-q^{-\alpha N}). \ee
Then equations $(S^\pm)^Nf(t)=\alpha^\pm f(t)$ take the following
simple form:
$$
(S^+)^N f(t)=(q-q^{-1})^{-N}q^{N t}(q^{\alpha N}f(t-N)-q^{-\alpha
N}f(t+N),
$$
and
$$
(S^-)^N f(t)=
(q-q^{-1})^{-N}q^{-Nt}(q^{-N\beta}f(t+N)-q^{N\beta}f(t-N)).
$$
The shift $t\to t\pm N$ in terms of variable $x$ means
multiplication of the argument by $q^{\pm N}=1$, i.e. identity
transformation. So we come to the following restriction for
functions realizing cyclic representation of the $s\ell_q(2)$
algebra in form of finite-difference equations: \be \label{de}
x^{\pm N}f(x)=const f(x). \ee It is well known \cite{mum} that
theta-functions possesses such kind quasi-periodicity property.
Theta-functions play the role of building blocks for functions
defining on Riemann surfaces, which is very similar to the role of
monomials $x^n$ for building functions on complex plane. However
in theory of Riemann surfaces dependence of theta-functions on the
second argument $\tau$ has crucial significance, in this context
the only meaning of second argument is that it provides
convergence of corresponding series.\\
It is reasonable to start the construction of representation after
little step back. $s\ell_q(2)$ algebra is closely related to the
Weyl algebra which is formed by two generators $X$ and $Z$: \be
 \label{wg}
 ZX=qXZ,
\ee their arbitrary powers and unity. In the space of functions
generators (\ref{wg}) can be realized in many ways. We mention two
of them: using of our finite-difference operators the generators
of Weyl algebra can be realized in most simple way as follows: \be
X=x,\qquad Z=q^{x\partial}. \ee Another realization in which we
are interested in related to the theory of theta-functions. Recall
\cite{mum}, that theta-function is the analytic, which determined
by its Fourier series: \be \theta(t,\tau)=\sum_{n=-\infty}^\infty
\exp{i\pi(n^2\tau+2nt)},\quad Im\tau >0. \ee Theta-functions with
characteristics are introduced using quasi-translation operators
$S_a$ and $T_b$: \be \label{tscr}
 S_bf(t)=f(t+b),\qquad
T_af(t)=e^{i\pi(a^2\tau+2at)}f(t+a\tau),
\ee
which form following algebra:
$$
S_aS_b=S_{a+b},\quad T_aT_b=T_{a+b},\quad S_bT_a=e^{2\pi
iab}T_aS_b.
$$
Then theta-functions with characteristics $a$, $b$ is defined as
result of action of these operators on usual theta-function: \be
\theta_{a,b}(t,\tau)\equiv
S_bT_a\theta(t,\tau)=\sum_{n=-\infty}^{+\infty}e^{
i\pi[(n+a)^2\tau+2(n+a)(t+b)]}. \ee Easy to see that
quasi-translations operators (\ref{tscr}) form the Weyl algebra if
one sets $q=e^{2i\pi ab}$. Hence theta-functions with
characteristics form the space of representation of the Weyl
algebra. In the case when parameter $q$ is equal to the root of
unity, the generators of Weyl algebra become nilpotent while
algebra itself becomes cyclic. In matrix form in this case
generator $Z$ is given by diagonal square matrix with powers $q^k$
$k=0,1,2,\ldots N-1$ while $X$ is square matrix with unities over
diagonal and in lower left corner. The representation space is
given by the set of $N$ theta-functions with characteristics $b=0$
and $a_k=\frac kN$. Indeed setting: \be \label{xz} Z=S_{\frac
1N},\qquad X=T_{\frac 1N} \ee one can check that these operators
obey to the Weyl algebra (\ref{wg}) with $q=e^{2\pi i/N}$ and act
on mentioned set of theta-functions: \be \label{theta}
\theta_k(t,\tau)\equiv (T_{\frac
1N})^k\theta(t,\tau)=\sum_{n=-\infty}^{+\infty}e^{ i\pi[(n+\frac
kN )^2\tau+2(n+\frac kN)t]}, \ee in following way:
$$
Z\theta_k(z,\tau)=e^{\frac{2\pi
i}{N}}\theta_k(t,\tau)=q\theta_k(t,\tau),\qquad
X\theta_k(z,\tau)=\theta_{k+1}(t,\tau).
$$
Choosing basis of representation as follows:
\be
\label{thetak}
\theta_{k}(x|\tau)=\sum_{n=-\infty}^\infty
e^{i\pi\tau n^2}x^{Nn+k},
\ee
one can establish relations:
$$
\theta_k(q^{\pm 1}x|\tau)=\sum_{n=-\infty}^\infty e^{i\pi\tau n^2}
x^{Nn+k}q^{\pm(Nn+k)}=q^{\pm k}\theta_k,
$$
$$
x^{\pm 1}\theta_k(x|\tau)=\sum_{n=-\infty}^\infty e^{i\pi\tau n^2}
x^{Nn+k\pm 1}=\theta_{k\pm 1}(x|\tau).
$$
It follows now that spin operators act on basis as:
$$
S^-\theta_k=q^{-\frac\lambda 2}[k-\beta]_q\theta_{k-1},
$$
\be\label{sthetak} q^S\theta_k+q^{k-\frac
12(\alpha+\beta)}\theta_k \ee
$$
S^+\theta_k=q^{\frac\lambda 2}[\alpha-k]_q\theta_{k+1}.
$$
It can be checked that basis theta-functions still unchanged upon
multiplication by $x^{\pm N}$. This is because being multiplied to
power function theta-function divides its power index by module
$N$.
\section{Tensor product of cyclic representations.}
In order to determine an action of the $R$-operator it is
necessary to realize tensor product of two representations on
which Casimir operators corresponding to $N$'s power of operators
(\ref{su}), (\ref{barsu}) would multiple of unit operator. The
tensor products of finite-dimensional irreducible representations
of $s\ell_q(2)$ at roots of unity were studied and decomposed into
direct sum of irreducible representations by D.Arnaudon \cite{ar}.
However in present context we are interested in a little bit
different approach. Let us define basis elements in this space as
follows: \be \label{2theta}
\theta_{k_1,k_2}=\sum_{n_i=-\infty}^\infty e^{i\pi
n_i\tau^{ij}n_j}x_1^{Nn_1+k_1}x_2^{Nn_2+k_2}, \ee here symmetric
matrix $\tau_{ij}$ is defined such that its imaginary part sets a
negative defined quadratic form. The generators $Z_i$ and $X_i$ do
not lead out from this set of functions:
$$
Z_i\theta_{k_1,k_2}=q^{k_i}\theta_{k_1,k_2},\quad
X_1\theta_{k_1,k_2}=\theta_{k_1+1,k_2},\quad
X_2\theta_{k_1,k_2}=\theta_{k_1,k_2+1}.
$$
In fact, as it was shown by V. Bazhanov and Yu. Stroganov
\cite{bs}, The form of L-operator (\ref{L}) is not the most
general one when deformation parameter is given by the root of
unity $q=e^{\frac{2\pi in}N}$, where $N$ is primary number.
However their extension, which depends on six free parameters was
mainly intended to describe the Chiral Potts Model in the approach
of Yang-Baxter equation do not leads to any significant changes in
present context. The statement of Bazhanov and Stroganov consists
of following: the Lax operator (\ref{L}) with finite-difference
operators subjected to Bogolyubov transformation for positive- and
negative-frequency operators shifting additive variable $t$ in
positive and negative direction with arbitrary coefficients also
obeys to Yang-Baxter equation with standard $r$-matrix and $q$
given by a root of unity. So using of this more complicated Lax
operator leads only to redefinition of eigenvalues of Casimir
operators.\\
It is easy to see that our basis elements are eigenstates of new
Casimirs. Indeed, one has
$$
\!\!\! S_u^-\theta_{k_1,k_2}\!=\! q^{\frac
12(u\!-\!\lambda_1\!-\!\alpha_2\!-\!\beta_2)\!+\!k_2}[k_1\!-\!\beta_1]_q
\theta_{k_1\!-\! 1,k_2}+q^{\frac 12(\alpha_1\!+\! \beta_1\!-\!
u\!-\!\lambda_2)\!-\! k_1}[k_2\!-\!\beta_2]_q\theta_{k_1,k_2\!-\!
1},
$$
$$
S_u^+\theta_{k_1,k_2}=q^{\frac
12(\lambda_1-u-\alpha_2-\beta_2)+k_2}
[\alpha_1-k_1]_q\theta_{k_1+1,k_2}+q^{\frac
12(u+\alpha_1+\beta_1+\lambda_2)-k_1}\theta_{k_1,k_2+1}.
$$
Then for their $N$'s powers one obtains using arguments similar to
those above: \be \label{suN} (S_u^-)^N=q^{N(\frac
u2+S_2)}(S_1^-)^N+q^{-N(\frac u2+S_1)}(S_2^-)^N, \ee
$$
(S_u^+)^N=q^{N(-\frac u2+S_2)}(S_1^+)^N+q^{N(\frac
u2-S_1)}(S_2^+)^N,
$$
from which desirable statement is easily deduced:
$$
(S_u^-)^N\theta_{k_1,k_2}=
$$
$$
=\left(q^{\frac
N2(u-\alpha_2-\beta_2-\lambda_1)}(q^{-N\beta_1}-q^{N\beta_1})+
q^{\frac
N2(-u+\alpha_1+\beta_1-\lambda_2)}(q^{-N\beta_2}-q^{N\beta_2})
\right)\theta_{k_1,k_2},
$$
and in similar manner for operators $S_u^+$ and $\bar S_u^\pm$.\\
So the set of $N^2$ functions (\ref{2theta}) realize the space of
tensor product of two representations. We are interested in
construction of the eigenvalues of $R$-operator. Again, like as
for general values of $q$ a crucial role play Yag-Baxter
relations, which express commutativity of the $R$-operator with
operators (\ref{su}) and (\ref{barsu}). Let us look for
eigenstates of the $R$-operator as a linear combinations of the
basis vectors (\ref{2theta})
$$
\varphi_m\equiv\sum_{k=0}^{N-1}a_k\theta_{m-k,k}\qquad\qquad
m=0,1,2\ldots N-1.
$$
The operator $S_u^-$ acts on these states as follows:
$$
S_u^-\varphi_m=\sum_{k=0}^{n-1}a_k\left( q^{k+\frac
12(u-\alpha_2-\beta_2-\lambda_1)}[m-k-\beta_1]_q
\theta_{m-1-k,k}+\right.
$$
$$
\left.+q^{k-m+\frac 12(\alpha_1+\beta_1-\lambda_2-u)}
[k-\beta_2]_q\theta_{m-k,k-1}\right).
$$
One can deduce that choosing: \be \label{ak}
a_{k+1}=a_kq^{u-2+\frac
12(\beta_1+\beta_2-\alpha_1+\lambda_2-\lambda_1)},\quad
a_{k}=a_0q^{k[u-2+\frac
12(\beta_1+\beta_2-\alpha_1+\lambda_2-\lambda_1)]}, \ee one will
obtain: \be \label{su-phi}
S_u^-\varphi_{m}(\alpha_i,\beta_i,\lambda_i|x_i)= q^{-1+\frac
12(u-\lambda_1+\beta_2-\alpha_2)}[m+1-\beta_1-\beta_2]_q
\varphi_{m-1}(\alpha_i,\beta_i, \lambda_i|x_i). \ee It is easy to
check, that operator $S_u^+$ shifts the same combination to up by
unity: \be
\label{su+phi}S_u^+\varphi_m(\alpha_i,\beta_i,\lambda_i|x_i)= \ee
$$
=q^{1-\frac 12(u-\lambda_1+\beta_2-\alpha_2)}[\alpha_1+\alpha_2+1-m]_q
q^{-1+\frac 12(u+\lambda_2\!+\beta_1-\alpha_1)} \varphi_{m+1}
(\alpha_i,\beta_i, \lambda_i|x_i).
$$
The similar formulae take place also for operators $\bar S_u^\pm$:
\be \label{bars-phi} \bar
S_u^-\bar\varphi_m(\alpha_i,\beta_i,\lambda_i | x_i,u)=q^{1-\frac
12(u+\lambda_1+\beta_2-\alpha_2)}[m+1-\beta_1-\beta_2]_q\bar
\varphi_{m-1}(\alpha_i,\beta_i,\lambda_i | x_i,u), \ee
$$
\bar S_u^+\bar\varphi_m(\alpha_i,\beta_i,\lambda_i|x_i,u)=q^{\frac
12(u+\lambda_1+\beta_2-\alpha_2)-1}[\alpha_1-\alpha_2+1-m]_q\bar
\varphi_{m+1}(\alpha_i,\beta_i,\lambda_i|x_i,u),
$$
where \be \label{barphi} \bar\varphi_m(\alpha_i,\beta_i,\lambda_i
|x_i,u)\equiv\sum_{k=0}^{N-1}q^{k[2-u+\frac
12(\alpha_1+\alpha_2-\beta_1-\beta_2+\lambda_2-\lambda_1)]}\theta_{m-k,k}.
\ee So these relations are compatible each to other and to
Yang-Baxter equations: \be \label{Rm}
R_m(u)=q^{2-u+\alpha_2-\beta_2-\lambda_1}R_{m-1}=
q^{m(2-u+\alpha_2-\beta_2-\lambda_1)}R_0(u), \ee if one sets
$$
R(u)\varphi_m(\alpha_i,\beta_i,\lambda_i | x_i,u)=R_m(u)
\bar\varphi_m(\alpha_i,\beta_i,\lambda_i | x_i,-u),
$$
and
$$
R(u)\bar\varphi_m(\alpha_i,\beta_i,\lambda_i | x_i,u)=R_m(u)
\varphi_m(\alpha_i,\beta_i,\lambda_i | x_i,-u).
$$
In fact the eigenvalues of the $R$-operator can be turn to unity
by redefinition of eigenstates $\varphi_m$ (by absorption of phase
multipliers).
\section{Acknowledgements}
This work is supported by Volkswagen Foundation of Germany, by
INTAS grants 00-561, 00-0390 and by Swiss SCOPE grant.
\section{Conclusion}
In this way realizing the generators of the $sl_q(2)$ algebra
generators in form of differential, more precisely
finite-difference operators (\ref{dqop}) or (\ref{fdot}) it is
appears to be possible to construct the functional representation
(\ref{fk}), which allows to deal with representations of any
dimension and to construct tensor product of such representations
and define the universal $R$-operator on this product. Also it is
possible to find eigenstates of the $R$-operator in that space and
corresponding eigenvalues for general and for exceptional values
of deformation parameter. Presented method works both for
finite-dimensional both for infinite-dimensional representations.
In particular it is applicable to the cyclic representations at
roots of unity, which have no classic analogs.
\section{Appendix}
In this appendix we show how obtained formulae can be used to
construct $R$-matrix for given representations. For general values
of deformation parameter the representations are given by the same
spaces $\mathbb{C}^{2\ell+1}$, as in non-deformed case. So one has
for $\ell=\frac 12$:
$$
S^+=x\frac{q^{1-x\partial}-q^{x\partial-1}}{q-q^{-1}}=
\left(\begin{array}{cc}0&1\\0&0\end{array}\right),\quad S^-=\frac
1x\frac{q^{x\partial}-q^{-x\partial}}{q-q^{-1}}=\left(
\begin{array}{cc}0&0\\1&0\end{array}\right),
$$
$$
S=x\partial-1/2=1/2\left(
\begin{array}{cc}1&0\\0&-1\end{array}\right),\quad
q^{aS}=\left(\begin{array}{cc}q^{\frac a2}&0\\0&q^{-\frac
a2}\end{array} \right)
$$
normalized vectors are:
$$
\left(\begin{array}{cc} 0\\1\end{array}\right)=1, \qquad
\left(\begin{array}{cc} 1\\0\end{array}\right)=x.
$$
The tensor product of such representations is built to be:
$$
1=\left(\begin{array}{cccc}0\\0\\0\\1\end{array}\right),\qquad
x_1=\left(\begin{array}{cccc}0\\1\\0\\0\end{array}\right),\qquad
x_2=\left(\begin{array}{cccc}0\\0\\1\\0\end{array}\right),\qquad
x_1x_2=\left(\begin{array}{cccc}1\\0\\0\\0\end{array}\right),
$$
the spin operators are:
$$
S_u^+=S_1^+\otimes q^{-\frac u2+S_2}+q^{\frac u2-S_1}\otimes
S_2^+=
\left(\begin{array}{cccc}0&q^{\frac{u-1}{2}}&q^{\frac{1-u}{2}}&0\\
0&0&0&q^{-\frac{u+1}{2}}\\0&0&0&q^{\frac{u+1}{2}}\\0&0&0&0
\end{array}\right)
$$
$$
S_u^-=\left(\begin{array}{cccc}0&0&0&0\\
q^{-\frac{u+1}{2}}&0&0&0\\q^{\frac{u+1}{2}}&0&0&0\\0&q^{\frac{u-1}{2}}
&q^{\frac{1-u}{2}}&0\end{array}\right),\quad
S=\left(\begin{array}{cccc}1&0&0&0\\
0&0&0&0\\0&0&0&0\\0&0&0&-1\end{array}\right),\quad
q^{aS}=\left(\begin{array}{cccc}q^a&0&0&0\\
0&1&0&0\\0&0&1&0\\0&0&0&q^{-a}\end{array}\right).
$$
Casimirs are calculated to be:
$$
{\bf{C}}_u=\left(\begin{array}{cccc}q+q^{-1}&0&0&0\\
0&q^{-1}&q^{-u}&0\\0&q^u&q&0\\0&0&0&q+q^{-1}\end{array}\right),\quad
{\bf{\bar C}}_u=\left(\begin{array}{cccc}q+q^{-1}&0&0&0\\
0&q&q^u&0\\0&q^{-u}&q^{-1}&0\\0&0&0&q+q^{-1}\end{array}\right).
$$
and have following eigenvalues:
$$
\varphi_0=\bar\varphi_0=\left(\begin{array}{cccc}0\\0\\0\\1
\end{array}\right),\;
\varphi_1(u)=\left(\begin{array}{cccc}0\\q^{\frac{1-u}{2}}\\-q^{\frac{u-1}{2}}
\\0\end{array}\right),\;
\varphi^1_0(u)=\left(\begin{array}{cccc}0\\q^{-\frac{u+1}{2}}\\
q^{\frac{u+1}{2}}\\0
\end{array}\right),\;
\varphi_0^2(u)=(q+q^{-1})\left(\begin{array}{cccc}1\\0\\0\\0
\end{array}\right).
$$
Over barred quantities are obtained by replacement $q\to q^{-1}$
and by simultaneous change of sign of argument $\varphi(-u)$.
Looking at the form of eigenvectors it is easy to guess that
$R(u)$-matrix has block-diagonal form:
$$
R(u)=\left(\begin{array}{cccc}g&0&0&0\\
0&a&b&0\\0&c&d&0\\0&0&0&h\end{array}\right).
$$
The recurrence relation gives:
$$
R_1=R_0\frac{[u-1]_q}{[u+1]_q}
$$
then relations $R(u)\bar\varphi_1(u)=R_1\varphi_1(-u)$ and
$R(u)\varphi_1(u)=\tilde R_1\bar\varphi_1(-u)$ lead to the
following conditions:
$$
(q^{u-1}-q^{1-u})b=(q-q^{-1})R_1=(q^{u-1}-q^{1-u})c,
$$
$$
(q^{u-1}-q^{1-u})a=(q^u-q^{-u})R_1=(q^{u-1}-q^{1-u})d,
$$
while $R(u)\varphi_0=R_0\varphi_0$ gives $g=h=R_0$. In this way:
$$
R(u)=\frac{R_1}{q^{u-1}-q^{1-u}}\left(\begin{array}{cccc}q^{u+1}-q^{-u-1}&0&0&0\\
0&q^u-q^{-u}&q-q^{-1}&0\\0&q-q^{-1}&q^u-q^{-u}&0\\0&0&0&
q^{u+1}-q^{-u-1}\end{array}\right),
$$
in accordance with the initial relation. This $R$-matrix could be
obtained in other way too:
$$
R(u)=[u+1]_q\left(\bar\varphi_0(-u)\times\varphi^T_0(u)+\frac{1}{q+q^{-1}}
\bar S^+_{-u}\bar\varphi_0(-u)\times\varphi^T_0(u)S^-_u+\right.
$$
$$
\left.\frac{1}{(q+q^{-1})^2} \bar
(S^+_{-u})^2\bar\varphi_0(-u)\times\varphi^T_0(u)(S^-_u)^2\right)-
\frac{[u-1]_q} {q+q^{-1}}\bar\varphi_1(-u)\times\varphi^T_1.
$$
here we omitted the factor: $\frac{R_0}{[u+1]_q}$. Consider now
the case $\ell_1=\frac 12$, $\ell_2=1$. Spin operators take form:
$$
S_2=x_2\partial_2-1=\left(\begin{array}{ccc}1&0&0\\
0&1&0\\0&0&-1\end{array}\right),\quad q^{aS}=
\left(\begin{array}{ccc}q^a&0&0\\
0&1&0\\0&0&q^{-a}\end{array}\right),
$$
$$
S^+=x_2\frac{q^{2-x_2\partial_2}-q^{x_2\partial_2-2}}{q-q^{-1}}=
\sqrt{q+q^{-1}}\left(\begin{array}{ccc}0&1&0\\
0&0&1\\0&0&0\end{array}\right),
$$
$$
S^-=\frac{1}{x_2}\frac{q^{x_2\partial_2}-q^{-x_2\partial_2}}{q-q^{-1}}=
\sqrt{q+q^{-1}}\left(\begin{array}{ccc}0&0&0\\
1&0&0\\0&1&0\end{array}\right).
$$
These act on tensor product according to formulae:
$$
S^+=q^{-\frac u2+S_2}S_1^++q^{\frac u2-S_1}S_2^+=
$$
$$
\left(\begin{array}{cccccc}0&q^{\frac
u2}\sqrt{1+q^{-2}}&0&q^{1-\frac u2}&&\\0&0&q^{\frac
u2}\sqrt{1+q^{-2}}&&q^{-\frac u2}&\\0&0&0&&&q^{-1-\frac
u2}\\&&&0&q^{\frac u2}\sqrt{1+q^2}&0\\&&&0&0&q^{\frac
u2}\sqrt{1+q^2}\\&&&0&0&0\end{array}\right),
$$
$$
S^-=q^{\frac u2+S_2}S_1^++q^{-\frac u2-S_1}S_2^-=
$$
$$
\left(\begin{array}{cccccc}0&0&0&&&\\q^{-\frac
u2}\sqrt{1+q^{-2}}&0&0&&&\\0&q^{-\frac u2}\sqrt{1+q^{-2}}&0&&&\\
q^{1+\frac u2}&&&0&0&0\\&q^{\frac u2}&&q^{-\frac
u2}\sqrt{1+q^2}&0&0\\&&q^{-1+\frac u2}&0&q^{-\frac
u2}\sqrt{1+q^2}&0\end{array}\right).
$$
The lowest weight vectors are determined to be:
$$
\varphi_0=\bar\varphi_0=\left(\begin{array}{cccccc}0\\0\\0\\0\\0\\1
\end{array}\right),\quad
\varphi_1(u)=\left(\begin{array}{cccccc}0\\0\\\frac{q^{-\frac
u2}}{\sqrt{q+q^{-1}}}\\0\\-\frac{q^{\frac u2}}
{\sqrt{q+q^{-1}}}\\0
\end{array}\right).
$$
Comparing with the non-deformed case one can see that matrix
structure of the representation does not changed and all
difference consist of the replacement of ordinary numbers to the
quantum ones:
$$ R^q_{\frac
12,1}=\left(\begin{array}{cccccc}g&0&0&0&&\\0&a&0&b
&&\\0&0&d&0&b&0\\0&b&0&d^\prime&0&0\\&&b&0
&a^\prime&0\\&&0&0&0&g\end{array}\right).
$$
The recurrence relation for present case takes form:
$$
R_1=\frac{[u-\frac 32]_q}{[u+\frac 32]_q}R_0,
$$
Taking into account that $\varphi_0$ and $\varphi_0^3$ are
constant vectors, i.e. are independent on $u$ and inert with
respect to the replacement $q$ to $q^{-1}$ one immediately
obtains:
$$
g=R_0.
$$
Acting by the $R$-matrix on the lowest weight vectors one obtains
the following set of equations:
$$
d-\frac{cq^{u-1}}{\sqrt{1+q^2}}=q^{-1}R_1,\qquad\quad
b-\frac{aq^{u-1}}{\sqrt{1+q^2}}=-\frac{q^uR_1}{\sqrt{1+q^{-2}}},
$$
$$
d-\frac{cq^{1-u}}{\sqrt{1+q^{-2}}}=qR_1,\quad\qquad
b-\frac{aq^{1-u}}{\sqrt{1+q^{-2}}}=-\frac{q^{-u}R_1}{\sqrt{1+q^2}},
$$
which leads to
$$
c=\frac{\sqrt{q+q^{-1}}R_1}{[u-\frac 32]_q}=c^\prime ,\quad
d=\sqrt{q+q^{-1}}R_1\frac{[u-\frac 12]_q}{[u-\frac 32]_q}, \quad
a=\sqrt{q+q^{-1}}R_1\frac{[u+\frac 12]_q}{[u-\frac 32]_q}.
$$
The remaining relations give:
$$
R^q_{\frac 12,1}(u)=\frac{R_1}{[u-\frac 32]_q} \left(
\begin{array}{cccccc}
[u+\frac 32]_q &&&0&0&0\\&[u+\frac
12]_q&&\sqrt{q+q^{-1}}&0&0\\&&[u-\frac 12]_q
&0&\sqrt{q+q^{-1}}&0\\0&\sqrt{q+q^{-1}}&0&[u-\frac 12]_q&&\\0&0&
\sqrt{q+q^{-1}}&&[u+\frac 12]_q&\\0&0&0&&&[u+\frac 32]_q
\end{array}\right).
$$
which is also coincides with non-deformed result under replacement
of ordinary numbers to the quantum ones.\\
Let us consider then more consistent case $\ell_1=\ell_2=1$. The
vectors $\varphi_N^m(u)$ are given by the following matrix
expressions:
$$
\varphi_0(u)=\left(\begin{array}{ccccccccc}0\\0\\0\\0\\0\\0\\0\\0\\1
\end{array}\right),\;\;
\varphi_0^4(u)=\left(\begin{array}{ccccccccc}1\\0\\0\\0\\0
\\0\\0\\0\\0\end{array}\right),\;\;
\varphi_0^1(u)=\left(\begin{array}{ccccccccc}0\\0\\0\\0\\0
\\q^{-\frac u2-1}\\0\\{\frac u2+1}\\0\end{array}\right),\;\;
\varphi_0^3(u)=\left(\begin{array}{ccccccccc}0\\q^{-\frac u2-1}\\0
\\q^{\frac u2+1}\\0\\0\\0\\0\\0\end{array}\right),
$$
$$
\varphi_1(u)=\left(\begin{array}{ccccccccc}0\\0\\0\\0\\0
\\q^{1-\frac u2}\\0\\-q^{\frac u2-1}\\0\end{array}\right),\quad
\varphi_1^2(u)=\left(\begin{array}{ccccccccc}0\\q^{1-\frac u2}\\0
\\-q^{\frac u2-1}\\0\\0\\0\\0\\0\end{array}\right),\quad
\varphi_1^1(u)=\left(\begin{array}{ccccccccc}0\\0\\q^{-u}\\0\\q-q^{-1}
\\0\\-q^{-u}\\0\\0\end{array}\right),
$$
$$
\varphi_2(u)=\left(\begin{array}{ccccccccc}0\\0\\q^{1-u}\\0\\-1
\\0\\q^{u-1}\\0\\0\end{array}\right),\quad
\varphi_0^2(u)=\left(\begin{array}{ccccccccc}
0\\0\\q^{-2-u}\\0\\q+q^{-1}\\0\\q^{u+2}\\0\\0\end{array}\right).
$$
$R$-matrix again has the same block form as in non-deformed case
as it can be seen from the form of its eigenvectors:
$$
R_1(u)=[u-2]_q[u+1]_q\frac{R_0(u)}{[u+1]_q[u+2]_q},\qquad
R_2(u)=[u-2]_q[u-1]_q\frac{R_0(u)}{[u+1]_q[u+2]_q}.
$$
From the condition $R(u)\varphi_0(u)=R_0(u)\bar\varphi_0(-u)$ one obtains:
$$
g=R_0(u).
$$
The same is true for $\varphi_0^4$. The next eigenvalue leads for
$\varphi_1(u)$ leads to the set of equations:
$$
q^{1-\frac u2}e-q^{\frac u2-1}f=q^{-1-\frac u2}R_1(u),\quad
q^{1-\frac u2}f-q^{\frac u2-1}e=-q^{1+\frac u2}R_1(u),
$$
and the same for $\varphi_1^2(u)$. One has:
$$
e=[u]_q[u+1]_q\frac{R_0(u)}{[u+1]_q[u+2]_q},\qquad
f=[2]_q[u+1]_q\frac{R_0(u)}{[u+1]_q[u+2]_q}.
$$
The equations for $\varphi_0^1(u)$ and $\varphi_0^3(u)$ lead to
the same values for $e$ and $f$. While equations for
$\varphi_2(u)$, $\varphi_1^1(u)$ and $\varphi_0^2(u)$ result to:
$$
bq^{1-u}-d+cq^{u-1}=q^{-u-1}R_2(u),
$$
$$
dq^{1-u}-a+dq^{u-1}=-R_2(u),
$$
$$
cq^{1-u}-d+bq^{u-1}=q^{u+1}R_2(u),
$$
$$
bq^{-u}+d(q-q^{-1})-cq^{u}=q^{-u}R_1(u),
$$
$$
dq^{-u}+a(q-q^{-1})-dq^{u}=-(q-q^{-1})R_1(u),
$$
$$
cq^{-u}+d(q-q^{-1})-bq^{u}=-q^uR_1(u),
$$
$$
bq^{-u-2}+d(q+q^{-1})+cq^{u}=q^{2-u}R_0(u),
$$
$$
dq^{-u-2}+a(q+q^{-1})+dq^{u}=(q+q^{-1})R_0(u),
$$
$$
cq^{-u-2}+d(q+q^{-1})+bq^{u}=q^{u-2}R_0(u).
$$
This set is consistent and has following solution:
$$
d=[u]_q[2]_q\frac{R_0(u)}{[u+1]_q[u+2]_q},\qquad
c=[2]_q\frac{R_0(u)}{[u+1]_q[u+2]_q},
$$
$$
b=[u]_q[u-1]_q\frac{R_0(u)}{[u+1]_q[u+2]_q},\qquad
a=([u]_q[u+1]_q+[2]_q)\frac{R_0(u)}{[u+1]_q[u+2]_q}.
$$
The final result has form:
\be
\label{RZ11} R(u)=
\ee
$$
\!\!\!\!\!
\left(\begin{array}{ccccccccc}\!\!\![u+1][u+2]\!\!\!&\!\!\!&\!\!\!&\!\!\!
&\!\!\!&\!\!\!&\!\!\!&\!\!\!&\!\!\!
\\\!\!\!&\!\!\![u][u+1]\!\!\!&\!\!\!&\!\!\![2][u+1]\!\!\!&\!\!\!
&\!\!\!&\!\!\!&\!\!\!&\!\!\!\\\!\!\!&\!\!\!&\!\!\![u][u-1]\!\!\!
&\!\!\!&\!\!\! [2][u]&\!\!\!&\!\!\![2]&\!\!\!&\!\!\!\\ \!\!\!
&\!\!\![2][u+1]\!\!\!&\!\!\!&\!\!\! [u][u+1]\!\!\!&\!\!\!
&\!\!\!&\!\!\!&\!\!\!&\!\!\!\\\!\!\!&\!\!\!&\!\!\!
[2][u]&\!\!\!&\!\!\![u][u+1]+[2]\!\!\!&\!\!\!&\!\!\!
[2][u]&\!\!\!&\!\!\!\\\!\!\!&\!\!\!&\!\!\!&\!\!\!&\!\!\!&\!\!\!
[u][u+1]\!\!\!&\!\!\!&\!\!\![2][u+1]\!\!\!&\!\!\!\\\!\!\!&\!\!\!&\!\!\!
[2]&\!\!\!&\!\!\![2][u]&\!\!\!&\!\!\![u][u-1]\!\!\!&\!\!\!&\!\!\!
\\\!\!\!&\!\!\!&\!\!\!&\!\!\!&\!\!\!&\!\!\![2][u+1]\!\!\!&\!\!\!
&\!\!\![u][u+1]\!\!\!&\!\!\!
\\\!\!\!&\!\!\!&\!\!\!&\!\!\!&\!\!\!&\!\!\!
&\!\!\!&\!\!\!&\!\!\![u+1][u+2]\!\!\!
\end{array}\right).
$$

\end{document}